\documentclass[twocolumn,showpacs,preprintnumbers,amsmath,amssymb]{revtex4}

\usepackage{graphicx}
\usepackage{dcolumn}
\usepackage{bm}

\begin{document}


\title{Sub-Natural-Linewidth Quantum Interference Features Observed in Photoassociation of a Thermal Gas}

\author{R. Dumke}
\author{J. D. Weinstein}
\author{M. Johanning}
\author{K. M. Jones}
\altaffiliation{permanent address: Physics Dept., Williams College, Williamstown, MA 01267}
\author{P. D. Lett}

\affiliation{Atomic Physics Division, National Institute of Standards and Technology, Gaithersburg, MD 20899-8424}

\date{\today}

\begin{abstract}
By driving photoassociation transitions we form electronically
excited molecules (Na$_2^*$) from ultra-cold (50-300 $\mu$K) Na
atoms. Using a second laser to drive transitions from the excited
state to a level in the molecular ground state, we are able to split
the photoassociation line and observe features with a width smaller than the
natural linewidth of the excited molecular state.   The quantum interference which gives rise to this effect is analogous to that which leads to electromagnetically induced transparency in three level atomic $\Lambda$
systems, but here one of the ground states is a pair of free atoms while the other is a bound molecule.  The linewidth is
limited primarily by the finite temperature of the atoms. 
\end{abstract}


\maketitle



\begin{figure}

\includegraphics[width=\columnwidth]{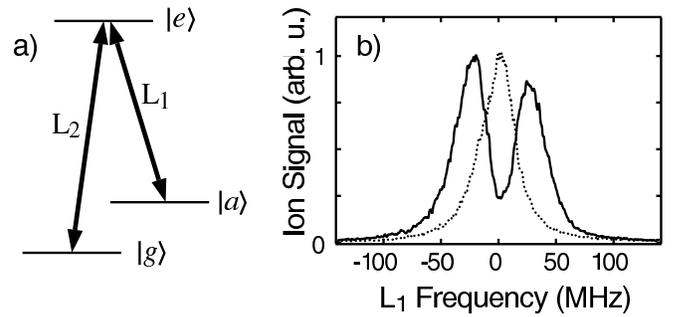}

\caption{\label{first} a) Atomic $\Lambda$ scheme often used to observe electromagnetically induced transparency. b) Photoassociation of 300~$\mu$K Na atoms to form molecules. The ion signal is a monitor of the population of electronically excited molecules and is shown as a function of the tuning of L$_1$,  without (dotted line) or with (black line) a second laser L$_2$ present. }

\end{figure}

When several coherent light waves interact with
multi-level atoms or molecules interesting phenomena emerge. In particular
electromagnetically induced transparency (EIT) \cite{EIT} and the closely associated  phenomena of ``slow'' light
 propagation \cite{Schmidt,SlowLight} have been extensively studied.  A convenient configuration for
observing EIT is an atomic ``$\Lambda$ system'' with two ground electronic state energy
levels each coupled by a coherent light field to a common excited level as
shown in Fig.~\ref{first}a.   Seemingly unrelated to the study of EIT, photoassociation
spectroscopy of ultracold atoms has also been an area of active
research \cite{Weiner,Jones}.  In a photoassociation transition, two free, colliding
atoms absorb a photon and become a bound, electronically excited
molecule.  If a second laser is introduced it is possible to have a level scheme reminiscent of the atomic $\Lambda$ system but with an important difference: one of the ``levels'' is replaced by a continuum of free colliding atom pairs while the other two levels are states of a bound molecule. Understanding which aspects of the behavior of the atomic system
carry over to the photoassociation system has been a topic of vigorous
discussion \cite{MackieReply} in the literature.  In this work we investigate whether it is possible to observe one of the characteristic features of EIT in the photoassociation system, namely is it possible to observe sub-natural-linewidth quantum interference features in the photoassociation of a thermal gas? 

EIT is commonly studied in alkali metal vapors.  The two ground states (Fig.~\ref{first}a),  $|g\rangle$ and $|a\rangle$,  are two hyperfine levels in the ground electronic state and $|e\rangle$ is an electronically excited state.  The
excited state has a short lifetime due to radiative
decay, while the ground states have much longer lifetimes, limited
by processes such as drift out of the experimental region, collisions or off-resonant scattering.  If laser L$_2$ is absent and laser L$_1$ is tuned over the appropriate frequency range, an absorption line due to $|a\rangle \rightarrow |e\rangle$ transitions is observed.  When L$_2$ is applied, and is tuned to be resonant with the $|g\rangle \leftrightarrow |e\rangle$ transition,  the absorption line produced by scanning L$_1$ can exhibit a narrow dip at the center.  The spectral width of this transparency window can be less than
the inverse lifetime of the excited state $|e\rangle$ \cite{Schmidt}, demonstrating that EIT is a
quantum interference effect. 

Figure \ref{first}b (dotted line) shows an example of a photoassociation resonance (Na + Na + $\gamma \rightarrow$ Na$_2^*$) observed in a
sample of 300~$\mu$K sodium atoms.  The height of the signal is
proportional to the number of molecules produced.  The solid line in Fig.~\ref{first}b shows the modified spectrum observed when a second light field is applied which couples the electronically excited
molecular state to a long lived vibrational level in the ground
electronic molecular potential.  The spectrum has a dip at line center where the photoassociation transition is suppressed.  In this particular case the dip is about the same width as the original photoassociation resonance.  Although the context is rather
different than the usual EIT experiment, the net effect is similar:
transitions due to one light field are suppressed by the introduction of a second light field.

 \begin{figure}
 \includegraphics[width=\columnwidth]{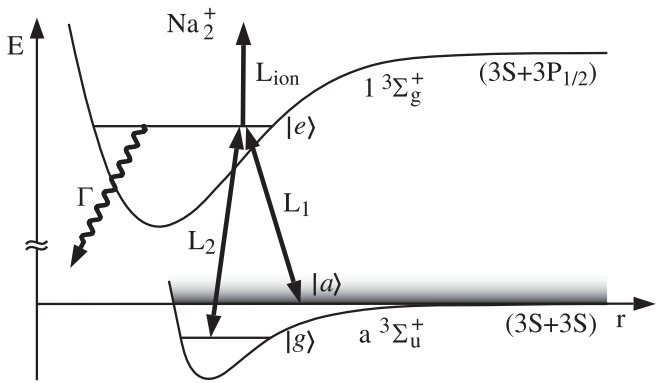}
\caption{\label{second} Sketch of the relevant Na$_2$ molecular potentials
and coupling lasers. The
photoassociation laser L$_1$ couples colliding atoms with energy $E$ (state $|a\rangle$) to form an excited-state molecule $|e\rangle$. The excited state
 is resonantly excited to an autoionizing state by
L$_\mathrm{ion}$, producing an ion signal proportional to the
excited state population.  L$_2$ couples the excited
molecular state to a single, long-lived rovibrational level $|g\rangle$ in a
ground state potential.}

\end{figure}

As shown in Fig.~\ref{second}, state $|a\rangle$ in the photoassociation system  is a pair of free colliding atoms with energy $E$ and is part of a continuum of such collision states.  The other two states, $|g\rangle$ and $|e\rangle$, are bound molecular states.  The continuum nature of the collision state makes for at least two important differences between this system and the atomic $\Lambda$ system.  One is that there is a spread in collision energies of order kT and thus a distribution of the population among different initial states $|a\rangle$.  Even if laser L$_1$ is on resonance for one particular collision energy it is off resonance for other collision energies. A second difference is that a pair of atoms with collision energy $E$, photoassociated to produce a molecule in state $|e\rangle$, can subsequently be photodissociated to produce two free atoms with an energy different from $E$, thus redistributing population within the continuum.  A third distinction between the two systems, which is not of central importance here, is that in the prototypical atomic system spontaneous emission from state $|e\rangle$ is primarily to one of the two ground states while in the molecular case spontaneous emission from $|e\rangle$ is likely to be to hot atoms or possibly bound molecular levels other than $|g\rangle$ and thus such spontaneous emission represents a loss from the system rather than a redistribution of population.

Two-color photoassociation of the type shown in Fig.~\ref{second} has been conducted primarily to locate bound levels $|g\rangle$ and the technique is often called ``Autler-Townes'' spectroscopy \cite{Jones,Araujo,Lisdat}.  In contrast to the data shown in Fig.~\ref{first}b, in such experiments L$_1$ is held fixed and L$_2$ is scanned, but the physics is the same.  The intensity of L$_2$ is usually kept high so that the width of the dip is larger than the natural linewidth of the excited state and, in this limit, the dip can be viewed as a consequence of the Autler-Townes splitting of the excited state.  The resulting two-color photoassociation
lineshapes have been well studied experimentally
\cite{Lisdat} and match nicely to calculations \cite{Bohn}
based on a scattering viewpoint discussed later. 

\begin{figure}
\includegraphics[width=\columnwidth]{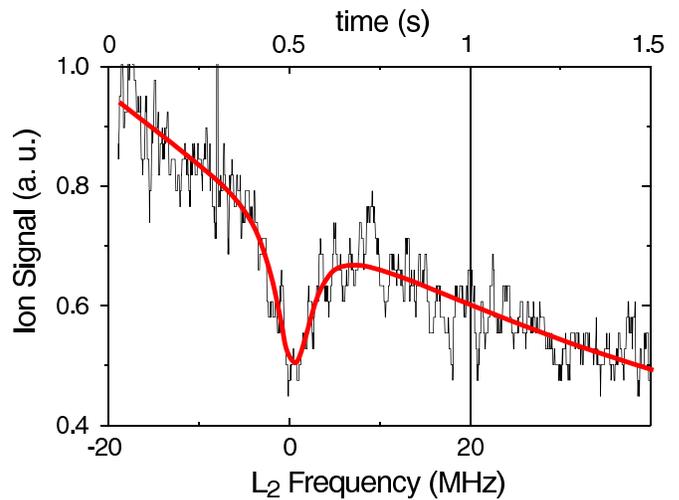}

\caption{\label{third} Photoassociation of sodium atoms held in a magnetic trap. L$_1$ is held fixed while L$_2$ is scanned over 40 MHz covering the $|e\rangle$ to $|g\rangle$ transition, stopping at the time indicated by the vertical line. The overall exponential decay of the signal is due to loss of atoms from the magnetic trap mostly by processes unrelated to photoassociation.  The dip in the ion signal indicates a suppression of population in level $|e\rangle$ when L$_2$ is on resonance for the $|e\rangle$ to $|g\rangle$ transition. The
superimposed Lorentzian curve has a linewidth (FWHM) of 5~MHz which is
narrower than the excited state linewidth of 17~MHz.}
\end{figure}

A typical experimental sequence starts with a standard
Zeeman-slower-fed six-beam dark-spot magneto-optical trap which produces a sample of $^{23}$Na atoms at about 300~$\mu$K. The
trapped atoms in the $F=1$, $m_F=-1$ state are transferred to a
magnetic quadrupole trap and evaporatively cooled to temperatures $\approx\!150~\mu$K, typically with densities of
$10^{12}$~cm$^{-3}$. We stop evaporative cooling at these elevated
temperatures to limit losses due to Majorana spin flips \cite{DarkSpot} which otherwise destroy the sample too quickly.  Under these conditions the sample decays with a 1/e lifetime of $\approx\!3$~s. By illuminating the atoms
with laser L$_1$ at 16403.89~cm$^{-1}$, we photoassociate free atoms to form bound excited molecules,  here in the $v=66$, $N=1$, $F=3$, $I=3$ level in the $1^3
\Sigma_g^+$ potential \cite{Araujo}.   The state $|g\rangle$ is chosen to be the $v=14$,
$R=0$, $I=3$, $F=2$ level of the $a^3 \Sigma_u^+$ potential.   It is located 6.659(0.015)~GHz below the
3S$(F=1)+3$S$(F=1)$ dissociation limit \cite{Araujo}.  L$_2$ is generated from L$_1$ by an electrooptic modulator so that their relative frequency difference is much better controlled than the nominal 1~MHz bandwidth of the cw dye laser which produces L$_1$.  When L$_2$ is generated in this way it is more convenient to hold laser L$_1$ fixed and scan L$_2$ in contrast to the method used to produce the spectrum in Fig.~\ref{first}b.

To monitor the population in the excited molecular state $|e\rangle$ an independent cw dye laser  L$_{\rm ion}$ at
17454.9~cm$^{-1}$ is used to drive transitions to an autoionizing level \cite{Amelink}.  The resulting Na$_2^+$ ions are detected by a microchannel plate. The excitation rate of molecules out of state $|e\rangle$ by L$_{\rm ion}$ is small compared to the spontaneous emission rate out of $|e\rangle$.
The linewidth of the autoionizing state
is greater than that of the other levels and greater than the Rabi frequencies in our
system.  Thus, this ionization pathway represents a weak perturbation on the system.

Figure \ref{third} shows a typical
measurement where  L$_2$ is scanned 40~MHz in 1 s. The atoms in the magnetic trap have a temperature of 50 $\mu$K (achieved by adiabatic lowering of the trap after evaporative cooling).
This data was taken with a power of 50~$\mu$W in L$_2$ and 45
mW in the photoassociation laser, L$_1$.  These Gaussian beams are focused to have a FWHM of $\approx$~40~$\mu$m at the trap.  The frequency of
L$_1$ is fixed to the peak of the one-color photoassociation resonance.  The ion signal decreases, with a 1/e lifetime of 1.2~s, as atoms are lost from the trap, primarily due to spin flips.  When L$_2$ is applied and scanned over the frequency of the $|g\rangle$ to $|e\rangle$ transition a dip is superimposed on the decay curve. The
measured linewidth of the dip, when fitted to a Lorentzian, is 5~MHz. The observed linewidth is smaller than the 17~MHz natural
linewidth of the excited state \cite{Araujo}. The dip is
centered at a laser difference frequency of 6.652~GHz which is approximately the binding energy of the target state $|g\rangle$.

\begin{figure}

\includegraphics[width=\columnwidth]{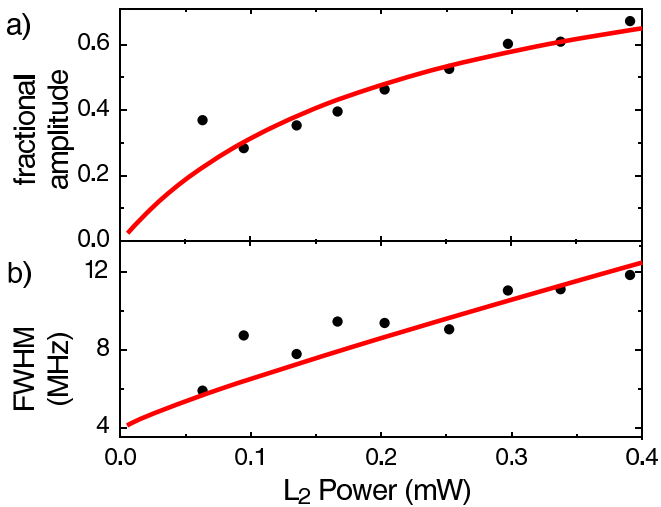}

\caption{\label{fourth}  a) Dip depth as a fraction of baseline ion signal and b)~FWHM of the observed dips, as a function of the power in L$_2$.  The fit lines are calculated from the theory of \cite{Bohn}.  Both the theoretical and experimental spectra are fit to a Lorentzian to extract an amplitude and a width.}
\end{figure}

Figure \ref{fourth} shows the behavior of the depth and width of the dip as a function of the power in L$_2$.  The width and amplitude of the lines where obtained by fitting the narrow spectral region containing the dip as a Lorentzian superimposed on a linear decay.

To compare experimental results to theory it is necessary to take into account the fact that the spot size of the laser beams is smaller than the cloud size of  $\approx100~\mu$m and thus our measured signals are an average over a range of Rabi frequencies.  To include this effect we assume a uniform atom density over the laser beam and calculate spectra using the scattering theory of \cite{Bohn}.  The natural linewidth of $|e\rangle$ is set to be 17~MHz and L$_1$ is assumed to be tuned to the peak of the one-color photoassociation line.   An amplitude and FWHM are extracted from the calculated spectra by fitting the dip with a Lorentzian as was done for the experimental data.  Two adjustments are made to fit the data.  A good fit is found for a temperature of 150~$\mu$K, consistent with independent measurements based on expansion of the atomic cloud which give values in the range of 100 - 180~$\mu$K for these trap conditions.  The one completely free parameter in the fit is the coefficient connecting the measured power of laser L$_2$ to the square of the Rabi frequency of the $|g\rangle \rightarrow |e\rangle$ transition at the center of the laser beam.  A value of  $780$~MHz$^2$/mW gives a good fit to the data.  Thus using reasonable input parameter values, the scattering theory gives a good fit to the data.  There is, at the level of the present measurements, no evidence for additional processes such as decay out of state $|g\rangle$.  The dominate contribution to the linewidth at low laser intensity is the thermal energy distribution of the atoms (at T~=~150~$\mu$K, $k_{\rm B} T/h \approx 3$~MHz).

\begin{figure}
\includegraphics[width=\columnwidth]{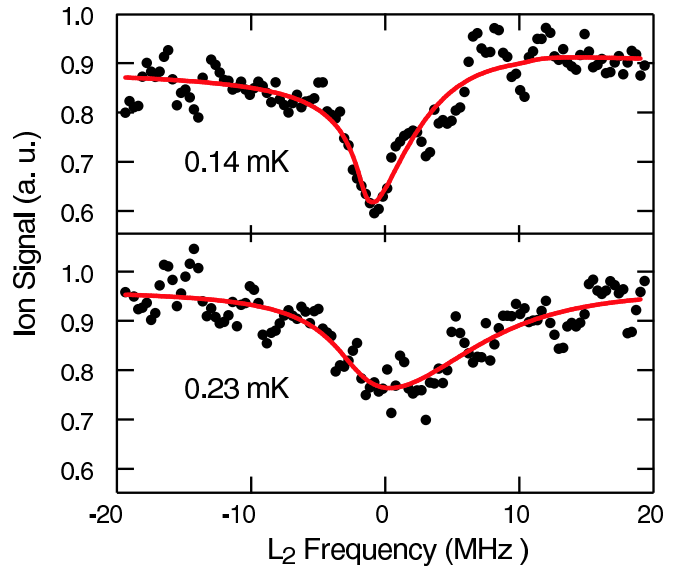}
\caption{ \label{fifth} Two spectra showing the effect of temperature.  The signal has been corrected for the overall decay of atom number.  The solid lines are calculated from the scattering theory of \cite{Bohn} averaged over the Gaussian distribution of Rabi frequencies in the experiment.  A Lorentzian fit to the 140~$\mu$K data yields a FWHM of 6~MHz, while the 230~$\mu$K data
is significantly broader with a FWHM of 12~MHz. }
\end{figure}

The strong role that temperature plays can be seen in Fig.~\ref{fifth}.  The solid curves are calculated from the scattering theory of \cite{Bohn} as before.  The temperatures were adjusted to fit the data and are consistent with independent measurements of the cloud temperature.

The data clearly shows that one can observe sub-natural-linewidths
in this system.  The detailed interpretation of this result depends on the
point of view one adopts to describe photoassociation.  Here we consider
two, one is the scattering view
exemplified by \cite{Bohn}, the other is the
quantum optics view, exemplified by
\cite{Mackie,Mackie2}.

In the scattering picture, photoassociation is a process which
converts colliding  atoms into detected product, in our case
Na$_2^+$ ions.  Within this framework one can ascribe the dips in the ion signal to quantum
interference between the multiple pathways leading from the initial
collision to the final product, i.e. between the paths $|a\rangle
\rightarrow |e\rangle  \rightarrow |ion\rangle$ and $|a\rangle
\rightarrow |e\rangle  \rightarrow |g\rangle \rightarrow |e\rangle
\rightarrow |ion\rangle$.  Viewed in this way the quantum interference which produces the dip is purely molecular and the role of photoassociation is to
provide input into this molecular interferometer.  The interferometer is tuned
to suppress ion production for a particular initial collision
energy.  Only for atoms at this one energy is the interference complete
and no ions produced.  The dip observed in the experiment is an average over
the thermal spread of collision energies, only a portion of which is
directly ``on resonance'' for the interferometer.   The lower the
temperature of the gas, the smaller the spread in collision energies
and hence the more nearly all of the flux is fed to the
interferometer on resonance. The scattering viewpoint leads to the equations which were used to fit the data in Figs.~\ref{fourth} and \ref{fifth} and provides a completely satisfactory description of the
observations.

In the quantum optics view of photoassociation, calculations are
done as if the system were contained in a finite-sized box and the
continuum limit is taken only at the end.  In this way the collision
continuum is replaced by a quasi-continuum of closely-spaced levels.
By focussing on one collision state at a time this view emphasizes
the connection with familiar bound-state systems, such as the three-level $\Lambda$ systems used to demonstrate EIT.  One way to describe the origin of the EIT dip in the $\Lambda$
system is to say that the system has been put into
a dark state, i.e. a coherent superposition of the two ground states $|a\rangle$ and $|g\rangle$
with the proper relative phase so that the system no longer absorbs
light.  Applying this particular view to the photoassociation system leads to the interpretation that we have created a rather unusual kind of 
dark state: a coherent superposition of two colliding atoms and a bound molecule.  The laser tunings are only appropriate to create
a dark state for one particular collision energy; for other
collision energies the superposition state is only ``grey."

In the present experiment the thermal spread of collision energies
sets a lower limit on the width of the dips we could observe.
If the thermal sample were replaced by a Bose-Einstein condensate (BEC), then much
narrower features should be observable.  In the scattering viewpoint there is nothing particularly special about the BEC, the main feature is that reducing the spread of incident
energies ensures that the incident flux is delivered to the
interferometer right on resonance.  In the quantum optics view of \cite{Mackie,Mackie2} a
BEC needs to be treated differently from a thermal gas.
The indistiguishibility of the atoms requires that all of them be
treated at once and leads to Bose enhancement factors \cite{Mackie2} which privilege the one initial
macroscopically-occupied level over other
levels in the collision continuum.

In the final stages of preparing this manuscript we learned that
this experiment has indeed been done using a BEC
instead of a thermal sample \cite{Winkler}.  The data presented in
the pre-print available to us shows a beautiful sharp dip, much
narrower than those we observed in a thermal gas.  While
quantitatively different, the results are qualitatively the same: both experiments demonstrate
sub-natural-linewidth features.  Our results show that observation of a sub-natural-linewidth dip does not require a coherent atom sample and thus, in and of itself, does not imply a macroscopic coherence between atoms and molecules.  Other types of experiments would be required to demonstrate such a coherence.

We thank Paul Julienne, Eite Tiesinga, and William Phillips
for useful discussions. R.D. thanks the Alexander von Humboldt Foundation for financial support.

\end{document}